# Low frequency electrodynamics in the mixed state of superconducting NbN and *a*-MoGe films using two-coil mutual inductance technique

Somak Basistha[1], Soumyajit Mandal, John Jesudasan, Vivas Bagwe, and Pratap Raychaudhuri

*Tata Institute of Fundamental Research, Homi Bhabha Road, Mumbai 400005*

**Abstract:**

We investigate the low-frequency electrodynamics in the vortex state of two type-II superconducting films, namely, a moderate-to-strongly pinned Niobium Nitride (NbN) and a very weakly pinned amorphous Molybdenum Germanium (*a*-MoGe). We employ a two-coil mutual inductance technique to extract the complex penetration depth, $\tilde{\lambda}$. The sample response is studied through the temperature variation of $\tilde{\lambda}$ in the mixed state, where we employ a model developed by Coffey and Clem (CC model) to extract the different vortex lattice (VL) parameters such as the restoring pinning force constant (Labusch parameter), VL drag coefficient and pinning potential barrier. We observe that a consistent description of the inductive and dissipative part of the response is only possible when we take the viscous drag on the vortices to be several orders of magnitudes larger than viscous drag estimated from the Bardeen-Stephen model.

---

[1] Email: somak.b.94@gmail.com

# 1. Introduction

When a type II superconductor is subjected to a magnetic field greater than its lower critical field $H_{c1}$, the field penetrates the superconductor in the form of quantised flux tubes each carrying a flux of, $\phi_0 = h/2e$ Wb, called vortices[1]. When an external current is passed through the superconductor, each vortex experiences a Lorentz force, $F_L = \phi_0 \mathbf{J} \times \hat{\mathbf{n}}$, where $\mathbf{J}$ is the current density and $\hat{\mathbf{n}}$ the unit vector along the magnetic field. For a perfectly clean superconductor, this results in the movement of the vortices, which gives rise to dissipation. This dissipation gives rise to an effective viscous force, $-\eta \dot{\mathbf{u}}$, where $\eta$ is the viscous drag coefficient and $\mathbf{u}$ is the position of the vortex. However, in a real superconductor, the inevitable presence of defects in the solid, which act as random pinning sites for the vortices, changes this scenario. This gives rise to a finite critical current density ($J_c$), for the onset of dissipation above which $F_L$ can overcome the pinning force. In contrast, the situation is more complicated when the superconductor is subjected to an oscillatory drive such as an oscillatory current or magnetic field. Here, even when $|\mathbf{J}| < J_c$, the vortex can undergo small oscillatory motion about the pinning potential minima giving rise to finite dissipation. Understanding the dynamics of vortices under a.c. excitation is therefore of paramount importance for technological applications[2,3,4,5,6,7].

In recent years, there is a renewed interest in the study of superconducting thin films owing to their technological applications. Compared to bulk superconductors, the 2D vortex lattice in thin films is much more fragile and susceptible to perturbations[8,9,10,11]. Consequently, thermal[12,13] and quantum[14,15] fluctuations can play a much more dominant role in these systems. Recently, a variety of vortex liquid states where true dissipationless transport is not seen even at vanishingly low currents have indeed been reported[16,17,18] in weakly pinned superconducting films made of conventional superconductors. On the other hand, strongly disordered superconducting films also show several unusual behaviour, such as magnetic field induced emergent inhomogeneity[19], magnetic field induced transition from a superconductor to an insulator[20] or a metal that retains spectroscopic signatures of superconducting correlations[21].

In this paper, we investigate the low frequency (tens of kHz) vortex dynamics of superconducting thin films of two different materials: NbN and amorphous $Mo_{0.7}Ge_{0.3}$ (*a*-MoGe). In terms of pinning these two films are in two opposite extremes; *a*-MoGe is a very weakly pinned superconductor, whereas NbN is moderate-to-strongly pinned. While several groups have studied the vortex state in superconducting thin films at microwave frequencies[22,23,24,25,26,27,28], the low frequency electrodynamic response at kHz frequencies has rarely been studied[29,30]. Here, we employ a two-coil mutual inductance technique[31,32,33,34] to measure the complex penetration depth, $\tilde{\lambda}$, in the mixed state of the superconductor and analyse the data using established theoretical models of vortex dynamics. An earlier study[35] on *a*-MoGe showed that while these models can explain the temperature and magnetic field dependence of the real part of $\tilde{\lambda}^{-2}$ (the inductive response), it failed to explain the imaginary part of $\tilde{\lambda}^{-2}$ (the dissipative response). In this paper, we show that a consistent description of both the inductive and dissipative response is obtained both in the strong and the weak pinning limit only when the value of $\eta$ is taken to be several orders of magnitude larger than the conventional estimate obtained from Bardeen-Stephen theory[36,37]. We discuss some possible origins of this large value of $\eta$.

## 2. Theoretical background

The dynamics of vortices under the influence of small oscillatory excitation is governed by three forces: (i) the viscous drag force, (ii) the restoring force due to the combined effect of pinning and inter-vortex interaction and (iii) the Lorentz force due to the current. One of the early models was proposed by Gittleman and Rosenblum[38] (GR) who considered the scenario for a single vortex, neglecting vortex mass term and thermal effects. Under these approximations the equation of motion of the vortex is given by:

$$\eta \dot{\boldsymbol{u}} + \alpha_L \boldsymbol{u} = \phi_0 \boldsymbol{J}^{ac} \times \hat{\boldsymbol{n}} \quad (1)$$

$\boldsymbol{J}^{ac}$ is the ac drive current density, $\hat{\boldsymbol{n}}$ is the unit vector along the vortex, $\eta$ is the viscous drag coefficient on the vortices in absence of pinning and flux creep, and $\alpha_L$ is the restoring force constant (Labusch parameter[39]) of the vortex. This model captures the vortex dynamics as long as the thermally activated motion of the vortices is slow compared to the excitation frequency. The pinning parameters, $\alpha_L$ and $\eta$ in this model are interpreted in a mean-field sense[40], where they incorporate both the effects of pinning potential and the vortex-vortex interactions. Assuming harmonic solution ($\sim e^{i\omega t}$) of the equation of motion (1), we get $\boldsymbol{u} = \phi_0 \frac{\boldsymbol{J}^{ac} \times \hat{\boldsymbol{n}}}{(\alpha_L + i\omega\eta)}$. Substituting this value in the London equation[41] we obtain[40]:

$$\boldsymbol{A} = -\mu_0 \lambda_L^2 \boldsymbol{J}^{ac} + \boldsymbol{u} \times \boldsymbol{B} = -\mu_0 \left( \lambda_L^2 + \frac{\phi_0 B}{\mu_0(\alpha_L + i\omega\eta)} \right) \boldsymbol{J}^{ac} = -\mu_0 \tilde{\lambda}^2 \boldsymbol{J}^{ac} \quad (2)$$

Here, eqn. (2) assumes a form similar to the usual London equation, where the London penetration depth[42] ($\lambda_L$) is replaced by the effective complex penetration depth, $\tilde{\lambda}$. The term $\frac{\phi_0 B}{\mu_0(\alpha_L + i\omega\eta)}$ captures the effect of the pinning force and viscous drag, and in the low frequency limit, $\omega \ll \alpha_L/\eta$, reduces to the square of the well-known Campbell penetration depth[43,44,45] given by, $\lambda_C = \left( \frac{\phi_0 B}{\mu_0 \alpha_L} \right)^{1/2}$. It is often useful to rewrite $\tilde{\lambda}^2$ in terms of the complex vortex resistivity, $\rho_v$:

$$\tilde{\lambda}^2 = \lambda_L^2 + \frac{\phi_0 B}{\mu_0(\alpha_L + i\omega\eta)} = \lambda_L^2 + \frac{\lambda_C^2}{(1 + i\omega\tau_0)} = \lambda_L^2 - \frac{i \rho_v}{\mu_0 \omega} \quad (3)$$

where $\tau_0 = \eta/\alpha_L$ is the vortex relaxation time and $\rho_v$ is expressed in terms of dc flux flow resistivity ($\rho_{ff}$) as[40,41]: $\rho_v = \rho_{ff}\frac{i\omega\tau_0}{1+i\omega\tau_0}$; $\rho_{ff} = \frac{B\phi_0}{\eta} = \frac{B}{B_{c2}}\rho_n$, where $\rho_n$ is the normal state resistivity.

Eqn. (2) does not take thermal effects into consideration. Since the pinning potential barriers are finite, at any finite temperature there is a finite probability for a vortex to hop over the barrier from one pinning minimum to the next. This process, called thermally activated flux flow (TAFF) or flux creep, has the effect of relaxing the restoring pinning force over characteristic time scale of vortex hop. This effect becomes particularly important when measurements are done at low frequencies. Different models have been proposed to account for TAFF. Brandt[46] proposed a phenomenological model where $\alpha_L$ was multiplied by an exponential relaxation such that, $\alpha_L \to \alpha_L(t) = \alpha_{L0}\exp(-\frac{t}{\tau_B})$, where $\tau_B = \frac{\eta}{\alpha_{L0}}\exp(\frac{U}{k_BT})$, $U$ being the effective pinning potential barrier and $k_B$ is the Boltzmann constant. A more sophisticated model was proposed by Coffey and Clem[47] (CC) who added a random $U$ dependent Langevin force to simulate the thermal motion and then solving it similar to that of a particle undergoing Brownian motion in a periodic potential[48,49]. Both models lead to a similar modification of eqn. (3) where the complex vortex resistivity takes the form[50],

$$\rho_v \to \rho_v^{TAFF} = \rho_{ff}\frac{\epsilon+i\omega\tau}{1+i\omega\tau} \quad (4)$$

where $\tau$ is the relaxation rate and $\epsilon$ is a dimensionless parameter which is a measure of the weight of the flux creep phenomenon. The expressions for $\tau$ and $\epsilon$ are somewhat different in the two models. Here, we take the CC expression:

$$\epsilon = \frac{1}{I_0^2(\nu)}, \tau = \tau_0\frac{I_0^2(\nu)-1}{I_1(\nu)I_0(\nu)} \quad (5)$$

where $\nu = \frac{U}{2k_BT}$ and $I_0$ and $I_1$ are the zeroth and first order modified Bessel functions of the first kind. Consequently, CC model has three temperature dependent parameters, $\alpha_L, \eta$ and $U$. Furthermore, the normal skin depth ($\delta_{nf}$) arising from the electrodynamic response of the normal electrons introduces an additional correction such that eqn. (3) is modified into:

$$\tilde{\lambda}^2 = \left(\lambda_L^2 - i\frac{\rho_v^{TAFF}}{\mu_0\omega}\right) \Big/ \left(1 + 2i\frac{\lambda_L^2}{\delta_{nf}^2}\right) \quad (6)$$

Here, $\delta_{nf}$ has the phenomenological variation of the form $\sim \left(\frac{\left(\frac{2\rho_n}{\mu_0\omega}\right)}{1-f(t,h)}\right)^{1/2}$ which is complementary to the variation of $\lambda_L$, where $\lambda_L^2(t,h) = \frac{\lambda_L^2}{f(t,h)}$, both of which comes from analysing a superconductor in the framework of the two-fluid model. Here $\rho_n$ is the normal state resistivity and $f(t,h) = (1-t^4)(1-h)$ [47], with $t = T/T_c$ and $h = H/H_{c2}(t)$. The main challenge in quantitatively fitting the CC model to experimental data, is finding the appropriate temperature dependence of the parameters. The temperature dependence of $\alpha_L$ and $U$ is non-universal and vary based on the nature of pinning in the material. However, one can get a crude estimate of $\alpha_L$ and $U$ from energetic considerations. Equating $U$ with the condensation energy inside the vortex core[41],

$$U \sim \frac{\mu_0 H_c^2}{2} \times (\pi\xi^2 d) \quad (7)$$

where $H_c$ is the thermodynamic critical field, $\xi$ is the coherence length and $d$ is the thickness of the sample. On the other hand, $\alpha_L$ can be estimated by equating the total condensation energy in the displaced core with the stored elastic energy: $\mu_0 H_c^2(\pi\xi^2 d) = \frac{1}{2}\alpha_L \xi^2 d$, which gives[51,52]:

$$\alpha_L(t) \sim \mu_0 H_c^2 \quad (8)$$

Using the known empirical dependences[41,53], $H_c(t = \frac{T}{T_c}) \sim (1-t^2)$ and $\xi(t) \sim \left(\frac{1+t^2}{1-t^2}\right)^{1/2}$ one obtains,

$$U(t) \sim U_0(1-t^2)(1+t^2) \quad (9)$$

$$\alpha_L(t) \sim \alpha_{L0}(1-t^2)^2 \quad (10)$$

Earlier works suggest that replacing the zero field $T_c$ in the expression of $t$ with $T_c(H)$[54] (namely the superconducting transition temperature in magnetic field), for the pinning parameters $\alpha_L$, $\eta$ and $U$ gives better agreement with data. In addition, it was suggested by Feigel'man et al.[55] and Koshelev et al.[56] that thermal fluctuations can cause a smearing of the potential, which gives an additional exponential multiplicative decay factor of the form $e^{-\left(\frac{T}{T_0}\right)}$. As we will show later, in $a$-MoGe this exponential factor is the dominant temperature dependence of these parameters. The temperature dependence of $\eta$ can be estimated from the Bardeen-Stephen formula [36], $\eta(t) = \frac{\mu_0 \phi_0 H_{c2}(t)}{\rho_n}$ ( where $\rho_n$ is the normal state resistivity) and therefore follows the temperature dependence of the upper critical field, $H_{c2} \sim H_{c2\,(0)} \left(\frac{1-t^2}{1+t^2}\right)$. Thus, $\eta$ follows the phenomenological temperature dependence,

$$\eta(t) = \eta_0 \left(\frac{1-t^2}{1+t^2}\right) \quad (11)$$

From the normal state resistivity (2.33 $\mu\Omega$-m for NbN and 1.51 $\mu\Omega$-m [57] for $a$-MoGe) and London penetration depth ($\lambda_L(0) \sim 360$ nm for NbN and 587 nm for $a$-MoGe) we can estimate the quantity $\left(\frac{2\lambda_L^2}{\delta_{nf}^2}\right)$ that appears in the denominator of eqn. (6). At our measurement frequency, this quantity is $\sim 10^{-7} - 10^{-8}$ and therefore can be neglected by setting the denominator to unity. The final form of the CC equation that we have used in our analysis is:

$$\tilde{\lambda}^{-2} = \lambda^{-2} + i\delta^{-2} = \left(\lambda_L^2 - i\frac{\rho_{ff}}{\mu_0\omega}\frac{\epsilon + i\omega\tau}{1+i\omega\tau}\right)^{-1}$$

$$= \left(\lambda_L^2 - \frac{\rho_{ff}}{\mu_0\omega\,(1+(\omega\tau)^2)}\left(\omega\tau(\epsilon-1) + i(\epsilon + (\omega\tau)^2)\right)\right)^{-1} \quad (12)$$

It can be seen that in absence of creep (where $U_0 \to \infty$) the CC model reduces to the GR model (eqn. 3).

## 3. Sample and Experimental Details

The samples under investigation are a 5 nm thick superconducting NbN thin film and a 20 nm thick superconducting $a$-MoGe thin film both grown on (100) MgO substrates using d.c. magnetron sputtering and pulsed laser deposition respectively. Details of sample growth and characterisation can be found in references [58], [59] and [57]. In terms of pinning strength these two samples lie on two ends of the spectrum: the NbN film is strongly pinned while the pinning in $a$-MoGe thin film is extremely weak. The contrasting pinning properties of the two samples can be seen from fig. 1 (a) and 1 (b) where we plot the magnetisation ($m$) as a function of an applied DC magnetic field ($H$) for the two films at the same reduced temperature (3.5 K for NbN and 2 K for $a$-MoGe) to facilitate comparison of pinning strength, measured using a SQUID-vibration sample magnetometer (SQUID-VSM). Both samples show hysteresis in $m$-$H$ typical of a type II superconductor. However, while for NbN the hysteresis loop is slightly open even at 1 kOe, for *a*-MoGe the loop closes at 25 Oe showing its extremely weakly pinned nature.

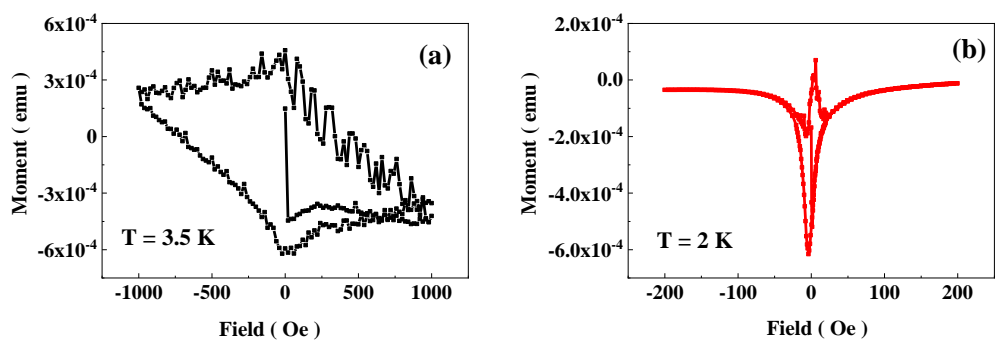

*Figure 1. (a) Variation of magnetic moment as a function of magnetic field (m – H) loop for a 5 nm thick NbN film grown on (100) oriented MgO at 3.5 K. (b) Variation of magnetic moment as a function of applied field for a 20 nm thick a-MoGe film grown on (100) oriented MgO at 2 K. The measurements are performed using a SQUID-VSM and have been done at nearly the same reduced temperature to compare the pinning strengths.*

To study the vortex dynamics, we measure the a.c. shielding response of the superconductor using the two coil mutual inductance technique. This method has been widely used to measure the London penetration of depth[31,33,34] of superconducting films. Here we use the same technique to measure $\tilde{\lambda}$ in the presence of an applied magnetic field. In this setup we sandwich an 8mm diameter superconducting thin film, grown using a shadow mask (left lower inset of fig. 2 (a)), in between a miniature quadrupolar primary coil and a dipolar secondary coil (right upper inset of fig. 2 (a)). Both the primary and secondary coils are wound on bobbins of 2mm diameter made out of Delrin. The quadrupolar primary coil consists of 15 turns clockwise in one half of the coil and another 15 turns counter-clockwise in the other half. The dipolar secondary coil consists of 120 turns wound in four layers. Both the coils are made of 50 $\mu$m diameter copper wires. An ac current ($I_{ac}$) with frequency ($f$) of 30 kHz is passed through the primary coil and the resulting in-phase ($V_{in}$) and out-of-phase voltage ($V_{out}$), in the secondary is measured using a lock-in amplifier. The radial profile of the magnetic field due to the current in the primary coil is shown in fig. 2 (a). The amplitude of $I_{ac} \sim 0.05$ mA corresponds to a peak magnetic field of 0.8 mOe. The complex mutual inductance ($M = M' + iM''$) between the two coils is given by $M'(M'') = V_{out}(V_{in})/(2\pi f I_{ac})$. The use of a quadrupolar drive coil is to ensure fast radial decay of the ac magnetic film on the film, which minimises the effect of edge effects[60,61,62] and geometric[63,64] barriers. For extracting the $\tilde{\lambda}^{-2} = \lambda^{-2} + i\delta^{-2}$ from $M$, we numerically solve the coupled Maxwell and London equations for the geometry of the coils and our sample using finite element analysis [31,33,34] and create a lookup table of $M = M' + iM$ for a range of values of $\lambda$ and $\delta$. $\tilde{\lambda}$ is then extracted by comparing the experimentally measured value of $M$ with the corresponding value in the lookup table. The measurement is performed in a ⁴He cryostat fitted with a 110 kOe superconducting solenoid and the magnetic field is always applied perpendicular to the film plane.

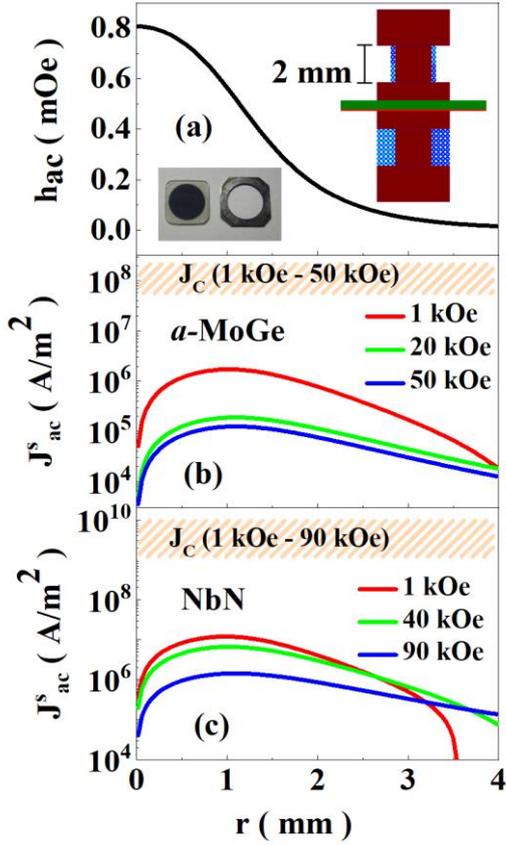

For analyzing the data using the CC model we need to ensure that the system is in the harmonic regime such that the oscillation of the vortices can be approximated by forced damped harmonic oscillators. This is satisfied by ensuring that the induced supercurrent density ($J_s^{ac}$) generated due to the ac drive applied through the quadrupolar coil, across the sample surface is much lower than the critical current density ($J_c$) of the sample. The induced current density on the superconducting film ($J_{film}^{ac}$), sandwiched between the quadrupolar drive coil and the dipolar pickup coil, is obtained from numerically solving the coupled Maxwell and London equations for the geometries involved and is given by[31,32],

*Figure 2. (a) Radial variation of ac field amplitude generated across the sample plane from passing 0.05 mA ac excitation through primary quadrupole coil (black); (inset: top right) schematic of the two-coil setup: quadrupole as drive coil (top) and dipole as pick-up coil (bottom) with sample sandwiched in between. The coil wire diameter ( 50 μm ) is drawn bigger than the actual for clarity; (inset: bottom left) 8 mm diameter sample grown on MgO substrate (left) using stainless steel mask (right). (b) Radial variation of the induced supercurrent density across the plane of 20 nm a-MoGe from passing 0.05 mA ac excitation through primary quadrupole coil at applied magnetic fields of 1 kOe, 20 kOe and 50 kOe. The shaded region in orange is the range of the critical current density from 1-50 kOe. (c) Radial variation of the induced supercurrent density across the plane of 5 nm NbN from passing 0.05 mA ac excitation through primary quadrupole coil at applied magnetic fields of 1 kOe, 40 kOe and 90 kOe. The shaded region in orange is the range of the critical current density from 1-90 kOe. All the current density values are calculated at 2 K.*

$$A_{film}(r) = A_{drive}(r) + \frac{\mu_0}{4\pi} \int d^3 r' \frac{J_{film}^{ac}(r')}{|r-r'|} \quad (13\text{ a})$$

$$\boldsymbol{A}_{film}(\boldsymbol{r}) = -\mu_0 \tilde{\lambda}^2 \boldsymbol{J}^{ac}_{film}(\boldsymbol{r}) \quad (13\ b)$$

where $\boldsymbol{A}_{film}(\boldsymbol{r})$ is the vector potential of the superconducting film, $\boldsymbol{A}_{drive}(\boldsymbol{r})$ is the vector potential associated with the drive coil and $J_s^{ac}$ is the real part of $J^{ac}_{film}$, the screening current density. Fig. 2 (b) and 2 (c) show the comparison between $J_c$ and $J_s^{ac}$ across the sample plane for the 20 nm $a$-MoGe and the 5 nm NbN respectively, at different magnetic fields. For NbN, we obtain the peak value of $J_s^{ac}$ ranging from $\sim 10^7 - 10^6$ A/m² in the field range $1 - 90$ kOe, while $J_c$ in the same field range is $10^{10} - 10^9$ A/m² [65]. For $a$-MoGe[66] we have $J_c \sim 10^8 - 10^7$ A/m² in the field range 200 Oe to 50 kOe, while the maximum $J_s^{ac}$ ranges from $10^6 - 10^5$ A/m². Thus, for both the samples we apply a subcritical drive, ensuring that we are well in the harmonic regime.

## 4. Results

### 4.1 Temperature variation of *M'*, *M"*, $\lambda^{-2}$ and $\delta^{-2}$

We summarize the screening response of the two superconductors as a function of temperature at various magnetic fields. Fig. 3 (a) shows the temperature variation of the screening response *(M')* for NbN. In absence of any applied DC magnetic field, we observe a clean and sharp transition for the 5 nm NbN film, showing a zerofield $T_c \sim 12$ K. This sharp single transition is a is characteristic of a clean and homogeneous film. As expected, when temperature is decreased, we see a monotonic decrease of *M'* below $T_c$ signifying an increase in the screening response of the superconductor. At the same time *M"* (fig. 3 (b)) shows a sharp dissipative peak. Unlike in zero field, the in-field *M"* does not go to zero even at temperatures well below the peak, showing that vortex induced dissipation continue to remain significant even at the lowest temperature. In fig. 3 (c) and 3 (d) we show $\lambda^{-2}$ and $\delta^{-2}$ extracted from the temperature variation of *M*. From the zerofield *M-T* we calculate the London penetration depth ($\lambda_L$) at 2 K ~ 360 nm. With increasing magnetic field $\lambda^{-2}$ decreases due to the increase in vortex contribution. $\delta^{-2}$ shows a broad peak near $T_c$ but remains finite at lower temperatures, showing that vortex dissipation remains significant at all temperatures. The very broad dissipative feature in this sample reflects the strongly disordered nature of the vortex lattice due to strong random pinning. For *a*-MoGe, *M'* (fig. 3 (e)) and $\lambda^{-2}$ (fig. 3 (g)) follow a similar monotonic trend to that of NbN. The zero-field transition is sharp with a $T_c$ of 7 K, pointing to the uniformness of the film. The contrasting feature in comparison to NbN, is that here at low fields *M'* ($\lambda^{-2}$) continue to decrease (increase) at low temperature instead of showing a tendency towards saturation. As we will show later this is related to the very weakly pinned nature of the vortex lattice of *a*-MoGe due to which fluctuation effects are larger. From the zerofield *M-T* the London penetration depth estimated at 2 K is 587 nm. Fig. 3 (f) and (h) show the temperature variation of *M"* and $\delta^{-2}$ at different magnetic fields. We observe that along with the main dissipation peak that appears just before $T_c$, there is a second broad minima that appears at lower magnetic fields (200 Oe – 1 kOe) which with increasing field gets gradually shifted out of the investigating temperature regime. This minima is however not of any special physical significance: it is not reflected in the temperature variation of $\delta^{-2}$ which decreases smoothly with increasing temperature and exhibits a single dissipative peak. In the next section we will quantitatively analyze the variation of $\lambda^{-2}$ and $\delta^{-2}$ with temperature and extract the different pinning parameters associated with the vortex dynamics.

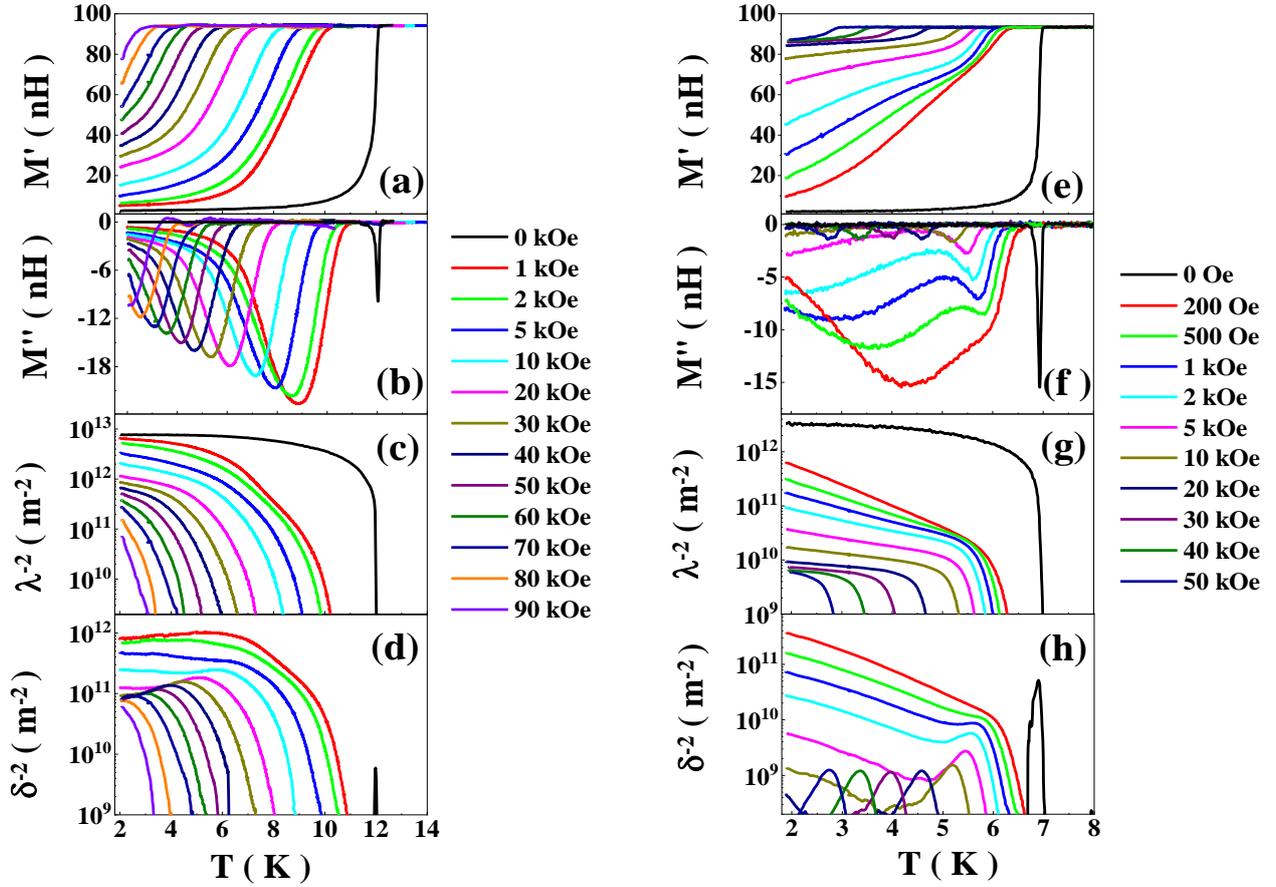

*Figure 3. (a) – (d) show the temperature variation of M' and M" and the temperature variation of the inverse square of the extracted penetration depth ( $\lambda^{-2}$ ) and the skin depth ( $\delta^{-2}$ ) at different magnetic fields respectively for a 5 nm thick superconducting NbN thin film. Figure 3. (e) – (h) show the temperature variation of M' and M" and the temperature variation of the inverse square of the extracted penetration depth ( $\lambda^{-2}$ ) and the skin depth ( $\delta^{-2}$ ) at different magnetic fields respectively for a 20 nm thick superconducting a-MoGe thin film. The measurements are done at an excitation amplitude of 0.05 mA and an operating frequency of 30 kHz.*

### 4.2.1 NbN

We first analyse the data at the lowest temperature of our measurements. Here thermal creep is negligible so we can use the GR model to extract the Labusch parameter ($\alpha_L$) and the viscosity coefficient ($\eta$) which is given as : $\alpha_L = \frac{\phi_0 B}{\mu_0} Re(\widetilde{\lambda_v}^{-2})$ and $\eta = \frac{\phi_0 B}{\mu_0 \omega} Im(\widetilde{\lambda_v}^{-2})$ where $\widetilde{\lambda_v}^{-2} = (\widetilde{\lambda}^2 - \lambda_L^2)$ is the square of the complex penetration depth comprising of only the vortex response. Fig. 4 (a) and (b) show the magnetic field variation of $\alpha_L$ and $\eta$ at 2 K for the 5 nm NbN, extracted using the GR model. $\alpha_L$ shows a weak increase at low fields giving rise to a broad maximum and then drops rapidly above 50 kOe where pinning potential starts getting shallower due to high magnetic fields. The shallow peak is a signature of collective pinning and will be discussed later. The most important point to note here is the magnitude of $\eta$ obtained. $\eta$ associated with the vortex motion is normally the Bardeen-Stephen (BS) flux flow viscosity which is typically of the order ~ $10^{-7} - 10^{-8}$ Ns/m$^2$. For NbN, using $\rho_n$ ~ 2.33 $\mu\Omega$-m and $H_{c2}$(2 K) ~ 16.08 Tesla [67] we obtain the BS viscosity at 2 K, $\eta_{BS}$ (2 K) ~ 1.43*10$^{-8}$ Ns/m$^2$. We observe that $\eta$ obtained from $Im(\widetilde{\lambda_v}^{-2})$ is nearly 5 orders of magnitude higher than the BS viscosity.

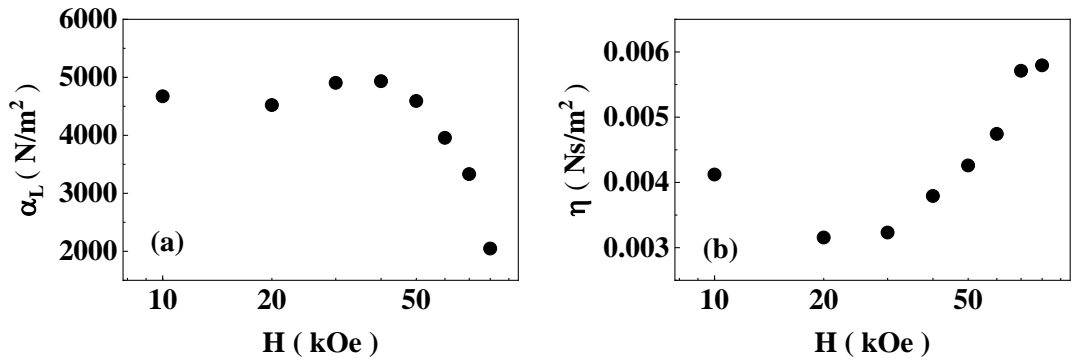

**Figure 4.** *(a) Variation of the Labusch parameter ($\alpha_L$) extracted using the GR model at 2 K, as a function of magnetic field for a 5 nm thick NbN film. (b) Variation of the coefficient of viscosity ($\eta$) extracted using the GR model at 2 K as a function of applied magnetic field for the same film.*

We now attempt to fit the temperature variation of $\widetilde{\lambda}^{-2}$. Here we use the CC model which incorporates the effect of flux creep. For $U_0$ and $\alpha_L$ we use the phenomenological temperature variations in eqn. (9) and eqn. (10) with the additional smearing factor discussed in sec. 2:

$\alpha_L \sim \alpha_{L0}(1-t^2)^2 e^{-T/T_0}$, $U \sim U_0(1-t^2)(1+t^2)e^{-T/T_0}$, where $t = \frac{T}{T_c(H)}$ is the scaled temperature For $\eta$ we use the phenomenological temperature variation in eqn. (11), $\eta \sim \eta_0 \left(\frac{1-t^2}{1+t^2}\right)$. We treat $U_0$ and $T_0$ as adjustable parameters and fix $\alpha_{L0}$ and $\eta_0$ in such a way, that their value at 2 K are as close as possible to the value extracted from the GR model. While fitting with eqn. (12), we tried to obtain the best fit for $\lambda^{-2} \equiv Re(\tilde{\lambda}^{-2})$ and then simulate $\delta^{-2} \equiv Im(\tilde{\lambda}^{-2})$ with the same set of parameters. The fitting procedure employed is discussed in details in the appendix. The fit is not perfect, which is unsurprising since the precise temperature dependence of various parameters is very difficult to predict, but the important point is that even within our approximations, the temperature variation of both $\lambda^{-2}$ and $\delta^{-2}$ are qualitatively very well captured. Fig. 5 shows the fit to the data using the same scheme for all magnetic field from 1 kOe to 90 kOe. The best fit parameters are given in Table 1. The simulated graphs capture the variation of $\lambda^{-2}$ and $\delta^{-2}$ for all fields except those below 5 kOe, where $\delta^{-2}$ shows an additional decrease at low temperatures. This could be because at these low fields the flux does not penetrate fully in the sample at low temperatures. Fig. 6 (a) - (d) shows the variation of $U, \alpha_L, \eta$ (at 2 K) and $T_0$ with magnetic field. $U$ follows a logarithmic decay with magnetic field going roughly as $\sim \frac{U_0}{k_B} \ln\left(H_0/H\right)$ from 20 kOe onwards, where $\frac{U_0}{k_B}$ ~ 21.6 K and $H_0$ ~ 164 kOe which is of the same order of magnitude as the $H_{c2}$ [68]. This form is predicted for collective pinning in a strongly disordered vortex lattice[69,70]. One can compare $U_0$ with its theoretically expected value[68], $U_0 = \frac{\phi_0^2 d}{64\pi^2 \mu_0 \lambda^2}$ in SI units. Using $\lambda = 360$ nm, we obtain $\frac{U_0}{k_B} \approx 15.04$ K which is of the same order of magnitude as obtained from the fit of $U$ vs $H$. It is also interesting to note that both the magnitude and magnetic field dependence of $U$ is similar to ones earlier extracted from d.c. transport measurements for samples with similar thickness[71,72]. Fig. 6 (b) shows the comparison between the magnetic field variation of $\alpha_L$ extracted from the GR and the CC model. In both cases $\alpha_L$ shows a weak increase at low fields giving rise to a broad maximum and then drops rapidly above 50 kOe where pinning potential starts getting shallower due to high magnetic fields. Within the collective pinning scenario, a maximum in $\alpha_L$ is expected due to enhanced pinning at the order-to-disorder transition of the vortex lattice (the "peak effect" [73,74]). Here the maximum is very broad, since the superconductor is strongly pinned[75] and the fully ordered vortex lattice is never realised. We observe that the magnetic field variation of $U$ and $\alpha_L$ point to the existence of collective pinning in the sample. The small difference between the values arise from the necessity to tune the $\alpha_L$ value for the CC model to get a good fit over the entire temperature range. However, the most striking observation in this study is the very large value of $\eta$, obtained both from GR and CC models which implies slow moving vortices and low dissipation. This is the most important finding of this work whose possible origin will be discussed later. Fig. 6 (c) shows the field variation of $\eta$ obtained from the two models. Similar to $\alpha_L$, the difference in values of $\eta$ between the GR and the CC model at high fields is due to tuning of $\eta$ in the CC model to get a good fit over the entire temperature regime. With increasing magnetic field, $\eta$ shows a shallow minima at 20 kOe and then monotonically increases upto 60 kOe. The trend beyond 60 kOe is not entirely reliable due to the difference between the GR and CC values. This non-monotonic behaviour of $\eta$ has not been understood yet. $T_0$ increases from 10 kOe to 20 kOe, after which there is a monotonic decrease till 80 kOe.

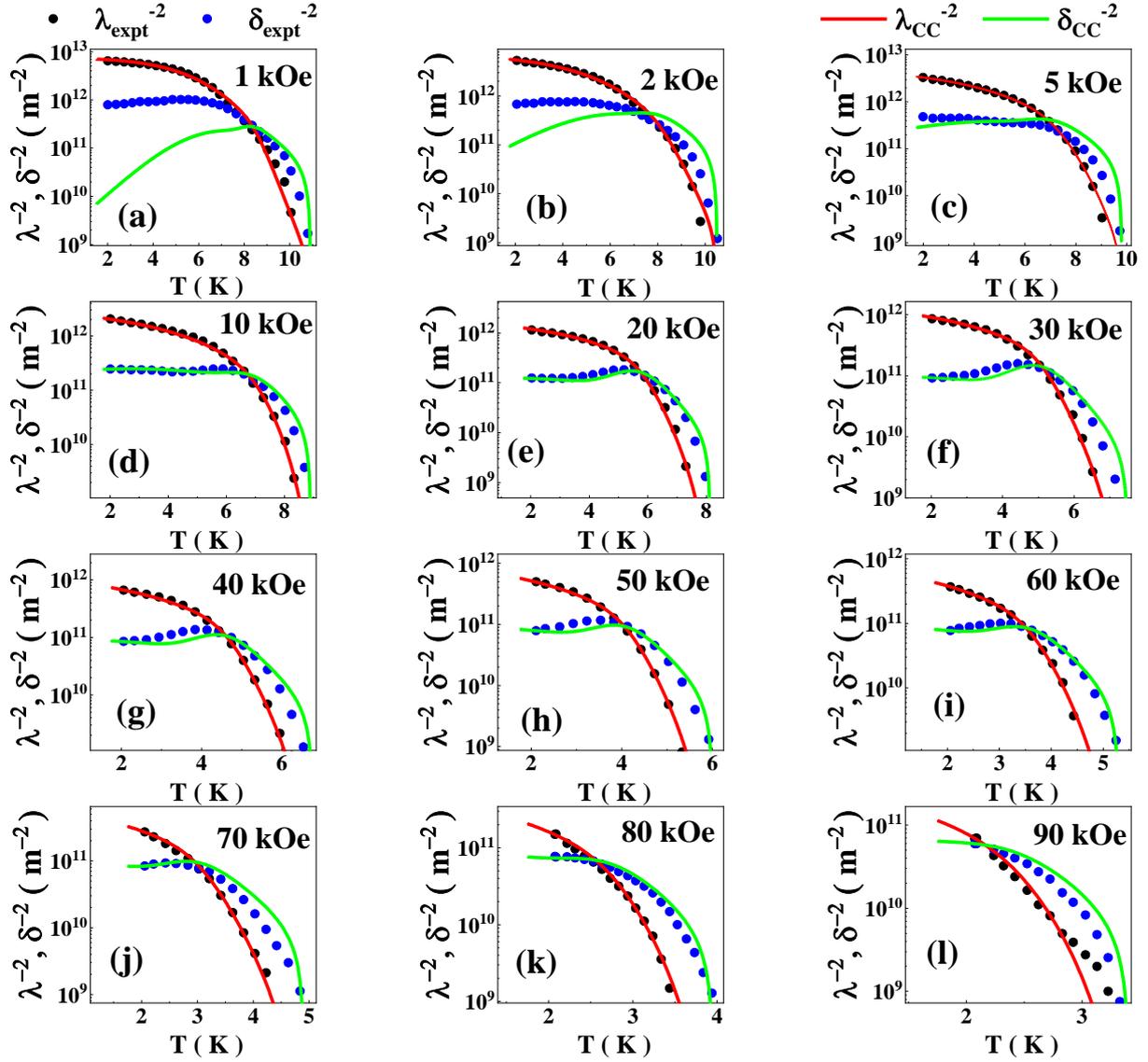

**Figure 5. (a) - (l)** *Temperature variation of experimentally measured $\lambda^{-2}$ and $\delta^{-2}$ for the 5 nm NbN thin film, at different magnetic fields from 1 kOe to 90 kOe along with the simulations performed using the Coffey-Clem model. From 10 kOe to 80 kOe the best-fit parameters are used for the simulation. The black and blue dots are the temperature variation of the experimentally measured $\lambda^{-2}$ and $\delta^{-2}$ respectively while the red and green solid lines are the simulated $\lambda^{-2}$ and $\delta^{-2}$.*

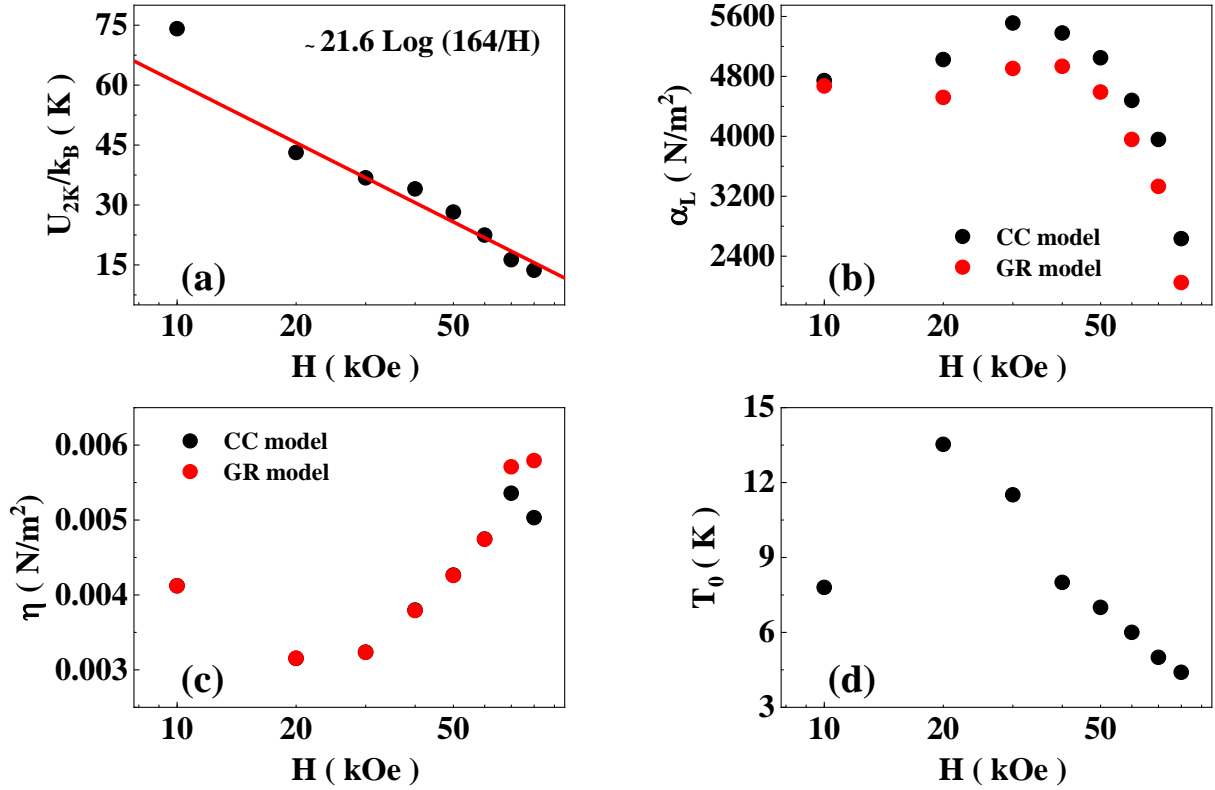

*Figure 6. (a) $U_{2K}/k_B$ vs H (black circles) which follows roughly a logarithmic decay (red line) in the field range 20 kOe – 80 kOe: $U_{2K}/k_B \sim (U_0/k_B)\log(164/H)$ where $(U_0/k_B) \sim 21.6$ K. (b) Magnetic field dependence of the pinning force constant, $\alpha_L$ calculated using the CC model (black circles) and the GR model (red circles). (c) Field variation of the viscosity coefficient, $\eta$, calculated using the CC model (black circles) and the GR model (red circles). (d) Field variation of the characteristic temperature $T_0$ (Used in the temperature dependence: $\alpha_L(T) = \alpha_{L0}(1-t^2)^2 e^{-T/T_0}$, $U(T) = U_0(1-t^2)(1+t^2)e^{-T/T_0}$ (black circles). The extracted parameters are for the 5 nm thick NbN thin film.*

**Table 1** *Best fit values of different parameters used for the simulation of the CC fits to the temperature variation of $\tilde{\lambda}^{-2}$ at different magnetic fields for 5nm NbN*

| Field (kOe) | $T_C(H)$ (K) | $T_0$ (K) | $U_0/k_B$ (K) | $\alpha_{L0}$ (Nm$^{-2}$) | $\eta_0$ (Nsm$^{-2}$) | $R^2$ for $\lambda^{-2}$ | $R^2$ for $\delta^{-2}$ |
|---|---|---|---|---|---|---|---|
| 10 | 8.9 | 7.8 | 96 | 6800 | 0.00456 | 0.998 | 0.943 |
| 20 | 8.12 | 13.53 | 52 | 6600 | 0.00356 | 0.994 | 0.942 |
| 30 | 7.5 | 11.51 | 44 | 7600 | 0.00373 | 0.998 | 0.877 |
| 40 | 6.75 | 8 | 44 | 8300 | 0.00453 | 0.992 | 0.891 |
| 50 | 6 | 7 | 38 | 8500 | 0.00533 | 0.998 | 0.934 |
| 60 | 5.3 | 6 | 32 | 8500 | 0.00632 | 0.996 | 0.963 |
| 70 | 4.9 | 5 | 25 | 8500 | 0.00750 | 0.993 | 0.846 |
| 80 | 3.95 | 4.4 | 23 | 7500 | 0.00850 | 0.990 | 0.955 |

### 4.2. 2. *a*-MoGe

We now focus on the 20 nm *a*-MoGe film. Similar to NbN, we analyse the data at 2 K where the effect of thermal creep is negligible and compute $\alpha_L$ and $\eta$ using the GR model. Fig. 7 (a) and 7 (b) shows the magnetic field variation of $\alpha_L$ and $\eta$ respectively. Here $\alpha_L$ shows a non-monotonic behaviour, where it displays a shallow minima around 10 kOe and then increases rapidly till 50 kOe. This is the signature of the onset of peak effect, which has been discussed in earlier papers [66, 35]. Here also we observe that the extracted $\eta$ which is typically associated with the Bardeen-Stephen (BS) flux flow viscosity, is 2 – 4 orders of magnitude larger than the BS estimate of the same, which can be calculated using $\rho_n \sim$ 1.51 $\mu\Omega$-m [57] and $H_{c2}$ (2 K) $\sim$ 10.8 Tesla [66] and comes out $\eta_{BS}$ (2 K) $\sim$ 1.47*10$^{-8}$ Ns/m$^2$. The recipe for fitting the temperature variation of $\tilde{\lambda}^{-2}$ using the CC model is similar to that used for NbN. However, as we have shown in an earlier work [35], in this case the exponential decay in the thermal smearing of the pinning landscape is more dominant than the prefactor. Therefore, we take the temperature variation of $U$ and $\alpha_L$ as, $\alpha_L \sim \alpha_{L0} e^{-T/T_0}$, $U \sim U_0 e^{-T/T_0}$. For the temperature dependence of $\eta$, we first use $\eta \sim \eta_0 \left(\frac{1-t^2}{1+t^2}\right)$, where $t = \frac{T}{T_c(H)}$ is the scaled temperature. Treating $U_0$ and $T_0$ as adjustable parameters, we fix $\alpha_{L0}$ and $\eta_0$ in such a way, that their value at 2 K are as close as possible to the value extracted from the GR model. With the above temperature variations, we see that although the fit to temperature variation of $\lambda^{-2}$ is well captured, however, the decrease in the measured value of $\delta^{-2}$ with increasing temperature is faster than the simulated value. The agreement becomes much better, if we incorporate the fluctuation factor ($e^{-T/T_0}$), in the expression of $\eta$, i.e. $\eta \sim \eta_0 \left(\frac{1-t^2}{1+t^2}\right) e^{-T/T_0}$.

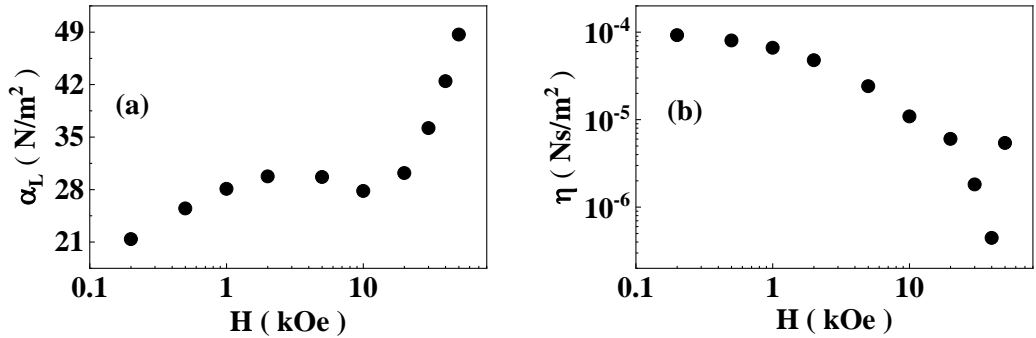

**Figure 7.** *(a) Variation of the Labusch parameter ($\alpha_L$) extracted using the GR model at 2 K, as a function of magnetic field for a 20 nm thick a-MoGe film. (b) Variation of the coefficient of viscosity ($\eta$) extracted using the GR model at 2 K as a function of applied magnetic field for the same film.*

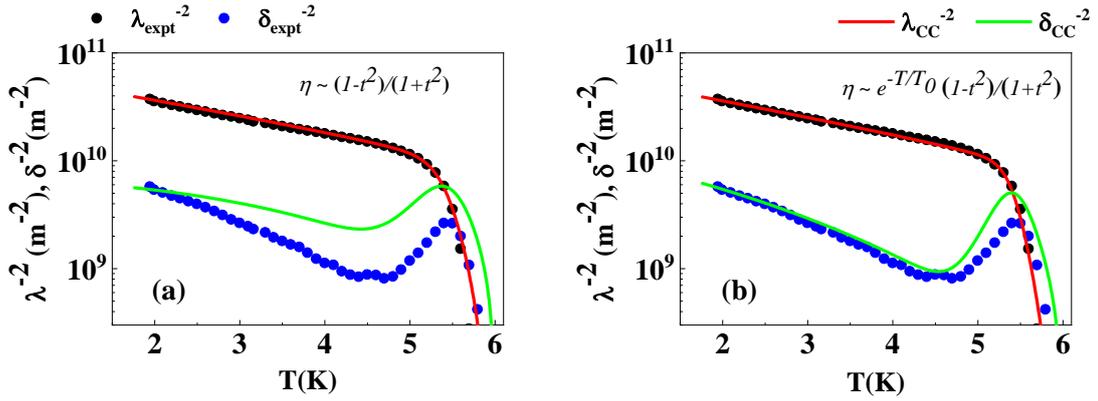

***Figure 8. (a) - (b)*** *Temperature variation of experimentally measured $\lambda^{-2}$ and $\delta^{-2}$ for a 20nm a-MoGe thin film, at 5 kOe, simulated with different models of the viscosity coefficient. The black and blue solid circles are the temperature variation of the experimentally measured $\lambda^{-2}$ and $\delta^{-2}$ respectively while the red and green solid lines are the simulations to $\lambda^{-2}$ and $\delta^{-2}$.* ***(a)*** *The simulation is done $\eta(T) = \eta_0 (1 - t^2)/(1 + t^2)$, the parameter values being : $\frac{U_0}{k_B} = 162\ K$, $\alpha_{L0} = 58\ Nm^{-2}$, $\eta_0 = 3 * 10^{-5}\ Nsm^{-2}$ and $T_0 = 3.2\ K$.* ***(b)*** *The simulation is done with $\eta(T) = \eta_0 e^{-\frac{T}{T_0}}(1 - t^2)/(1 + t^2)$, the parameter values being: $\frac{U_0}{k_B} = 230\ K$, $\alpha_{L0} = 60\ Nm^{-2}$, $\eta_0 = 6 * 10^{-5}\ Nsm^{-2}$ and $T_0 = 2.95\ K$.*

Fig. 8 shows the fit in both the cases. In Fig. 8 (b), the temperature variation of $\delta^{-2}$ is captured much better except for a small discrepancy in the magnitude of the dissipation peak. Physically, this exponential decay reflects a decrease in viscosity caused by the thermal fluctuation of vortices. We have fitted the temperature variation of $\tilde{\lambda}^{-2}$ by obtaining the best-fit for $\lambda^{-2} \equiv Re(\tilde{\lambda}^{-2})$ and then simulate $\delta^{-2} \equiv Im(\tilde{\lambda}^{-2})$ with the same parameters. Details about the fitting procedure is discussed in the appendix.

We now extend this procedure to fit the data for magnetic fields ranging from 200 Oe to 50 kOe (fig. 9). The fit parameters are given in Table 2. We observe that the qualitative behaviour of $\lambda^{-2}$ and $\delta^{-2}$ is reasonably captured at lower fields, but with increasing field beyond 5 kOe, the dissipative peak of $\delta^{-2}$ is overestimated in the simulation. This discrepancy results from the specific form for temperature dependence for various parameters assumed in our calculations. While we can improve the fit the by fine tuning the temperature dependences this introduces additional parameters and does not provide any significant physical understanding.

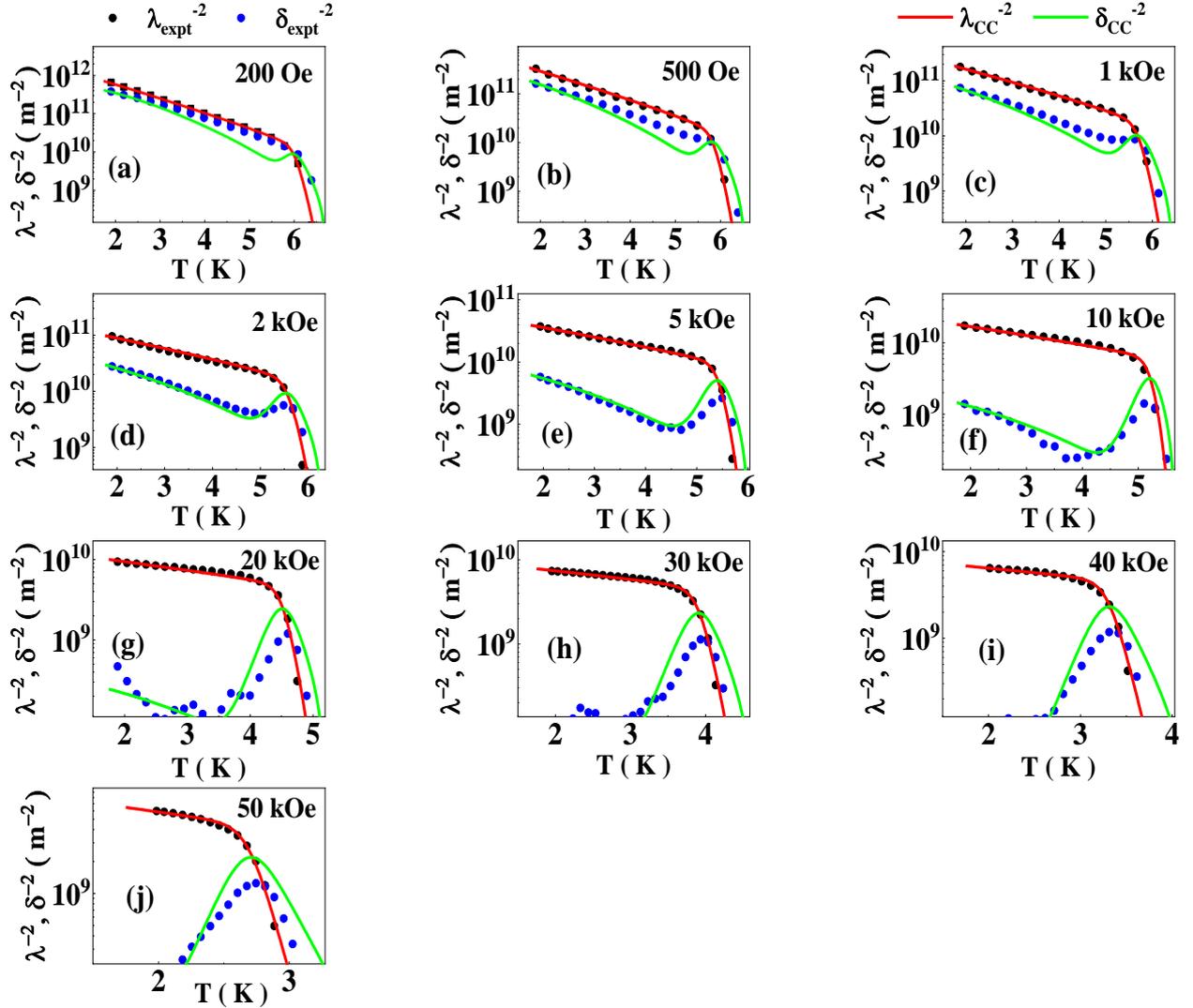

*Figure 9. (a) - (j)* Temperature variation of experimentally measured $\lambda^{-2}$ and $\delta^{-2}$ for the 20 nm a-MoGe thin film, at different magnetic fields from 200 Oe to 50 kOe along with the simulations performed using the Coffey-Clem model. The black and blue dots are the temperature variation of the experimentally measured $\lambda^{-2}$ and $\delta^{-2}$ respectively while the red and green solid lines are the simulated $\lambda^{-2}$ and $\delta^{-2}$.

In fig. 10 we explore the magnetic field variation of the parameters at 2 K. $U$ follows the generic power law decay with magnetic field with an exponent[76] $H^{-0.56}$. This kind of the power-law decay with exponent varying between 0.2 – 1 is generic of a wide range of weakly pinned superconductors[77,78,79,80,81,82]. $\alpha_L$ on the other hand, follows a non-monotonic behaviour; It displays a shallow minimum around 10 kOe and then increases up to 50 kOe. As discussed in an earlier paper[66], this increase is a signature of the onset of peak effect[73,74], which is a hallmark of collective pinning in a weakly pinned superconductor. The weak pinning nature is also reflected in the value of $\alpha_L$ which is two orders of magnitude smaller than NbN. $T_0$ increases monotonically with magnetic field, till 30 kOe, beyond which it shows a slight decrease. $\eta$ on the other hand, decreases by two orders of magnitude from 200 Oe to 50 kOe. (The behaviour observed beyond 30 kOe is not entirely reliable owing to the small range of the fit at these fields.) However, the absolute value of $\eta$ is still 2 orders of magnitude larger than the BS value even at the highest field. The difference in the values of $\alpha_L$ and $\eta$ between the GR and the CC model arises due to the need for minor adjustments of the parameter values during the fitting procedure using the CC model, such that the $R^2$ estimate is maximised for fits to both $\lambda^{-2}$ and $\delta^{-2}$.

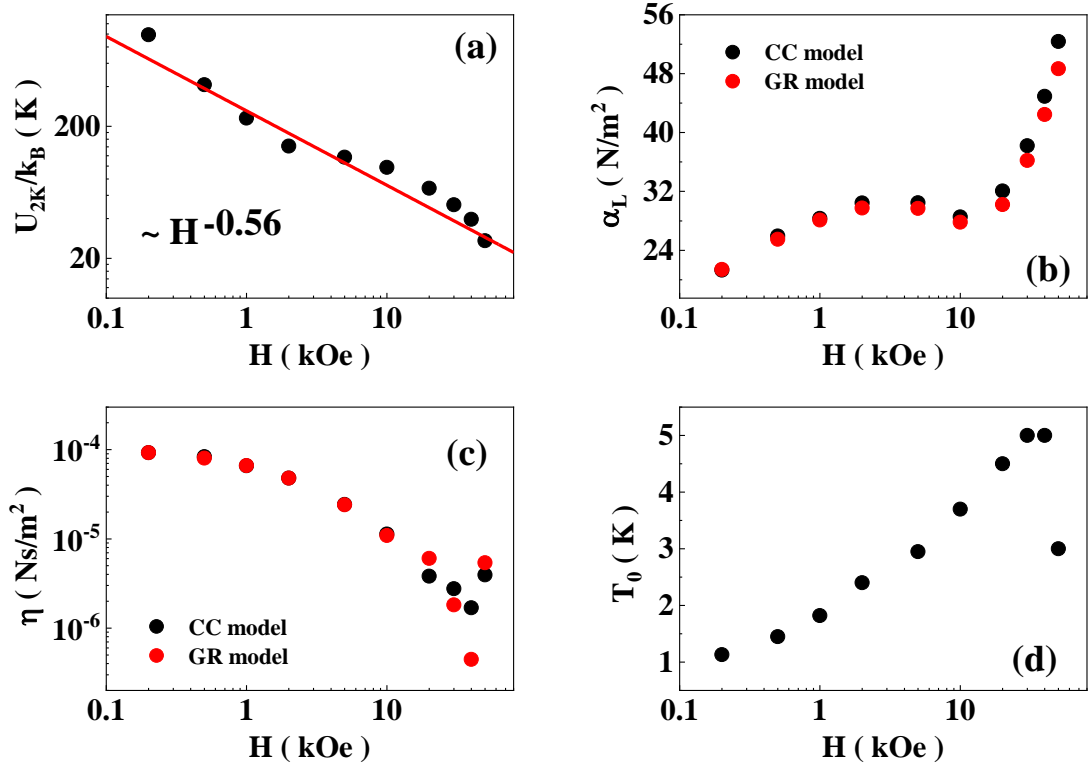

*Figure 10. (a) $U_{2K}/k_B$ vs H (black circles) which follows roughly a power law decay (red line): $U_{2K} \propto H^{-0.56}$ (b) Magnetic field dependence of the pinning force constant, $\alpha_L$ calculated using the CC model (black circles) and the GR model (red circles). (c) Field variation of the viscosity coefficient, $\eta$, calculated using the CC model (black circles) and the GR model (red circles). (d) Field variation of the characteristic temperature $T_0$ (Used in the temperature dependence of $\alpha_L(T) = \alpha_{L0}e^{-T/T_0}$, $U(T) = U_0 e^{-T/T_0}$ and $\eta(T) = \eta_0 e^{-T/T_0}(1-t^2)/(1+t^2)$. The extracted parameters are for the 20 nm thick a-MoGe thin film.*

**Table 2** *Values of different parameters used for the simulation of the CC fits to the temperature variation of $\tilde{\lambda}^{-2}$ at different magnetic fields for 20nm a-MoGe*

| Field (kOe) | $T_C(H)$ (K) | $T_0$ (K) | $U_0/k_B$ (K) | $\alpha_{L0}$ (Nm$^{-2}$) | $\eta_0$ (Nsm$^{-2}$) | $R^2$ for $\lambda^{-2}$ | $R^2$ for $\delta^{-2}$ |
|---|---|---|---|---|---|---|---|
| 0.2 | 6.7 | 1.13 | 5800 | 125 | 6.5*10$^{-4}$ | 0.997 | 0.857 |
| 0.5 | 6.55 | 1.45 | 1640 | 103 | 4*10$^{-4}$ | 0.997 | 0.889 |
| 1 | 6.45 | 1.82 | 692 | 85 | 2.4*10$^{-4}$ | 0.996 | 0.871 |
| 2 | 6.3 | 2.4 | 326 | 70 | 1.35*10$^{-4}$ | 0.992 | 0.848 |
| 5 | 6 | 2.95 | 230 | 60 | 6*10$^{-5}$ | 0.992 | 0.699 |
| 10 | 5.65 | 3.7 | 168 | 49 | 2.5*10$^{-5}$ | 0.956 | 0.520 |
| 20 | 5.2 | 6 | 106 | 50 | 8*10$^{-6}$ | 0.975 | 0.423 |
| 30 | 4.63 | 5 | 76 | 57 | 6*10$^{-6}$ | 0.986 | 0.458 |
| 40 | 4.2 | 5 | 59 | 67 | 4*10$^{-6}$ | 0.993 | 0.372 |
| 50 | 3.7 | 3 | 53 | 102 | 1.4*10$^{-5}$ | 0.985 | 0.621 |

## 5. Discussions

The central finding is this work is that when the vortices are subjected to a sub-critical low-frequency a.c. drive, $\eta$ is 2-5 orders of magnitude larger than the BS estimate which lies in the range ~ $10^{-7}$ to $10^{-8}$ Ns/m$^2$ for a majority of superconductors[50,83,84,85]. Even though the temperature dependence of various parameters is phenomenological, this central result is not sensitive to these details. On the other hand, $\eta$ extracted from microwave response of vortices in various superconductors is often consistent with the BS value[26,28,86,87]. Interestingly, at even higher frequencies in the THz frequency range $\eta$ values that are two orders of magnitude smaller that BS value was reported[88] in La$_{2-x}$Sr$_x$CuO$_4$. On the other hand, experiments probing slow vortex motion over larger time scales have reported large $\eta$. For example, low temperature vortex viscosity in YBa$_2$Cu$_3$O$_{7-\delta}$ measured with a bolometric technique[89] yielded an $\eta$ of the order $10^{-5}$ Ns/m$^2$ which exceeded the BS value by 2 orders of magnitude. Similarly, very large $\eta$ ( 0.1 – 284 Ns/m$^2$ ) have been reported from the early stages of magnetization relaxation in YBCO monocrystal when the external magnetic field is switched off and the magnetization relaxation is studied over time[90]. Bondarenko *et al*. obtained a viscosity of $10^{-5}$ Ns/m$^2$ during the study of locally frozen magnetic field in high temperature superconducting ceramic[91].

In BS model[36], $\eta$ is computed from the damping force that an isolated vortex experiences when it moves inside the superconductor. The vortices are treated as rigid cylinders with a normal core of radius $\xi$ with circulating supercurrent outside the core. Dissipation occurs through the viscous drag on the normal electrons both inside the core as well as the normal electrons which lies, in the normal to superconducting transition region, outside the core. In this model, several factors such as the interaction between vortices, interaction of vortices with pinning sites and the possibility of bending of the vortex line are ignored. Subsequently, several models have been proposed to explain deviations from BS value of $\eta$. Entanglement of the flux lines[92,93,94,95] due to their bending, in 3D superconductors have been proposed as a reason for large increase in $\eta$. The thickness of both films is in nm, while the vortex bending length scale is of the order of $\mu$m [96]. This implies that the vortices formed in both films are in the 2D limit, making vortex entanglement improbable. Another mechanism which can lead to the deviation of $\eta$ from the BS value, is the Hall viscosity that arises due to the Magnus force acting on the vortices[97]. The Magnus force arises due to the relative motion between the vortex core and the superelectrons and acts perpendicular to the motion of the vortices. The coefficient of the Magnus drag, which is the hall viscosity[85,98,99] ($\alpha_H$) adds to the BS drag as $\eta^* = \eta_{BS}\left(1 + \left(\frac{\alpha_H}{\eta_{BS}}\right)^2\right)$. Depending on the magnitude of $\alpha_H$ in comparison to $\eta_{BS}$ the Hall angle, $\left(\theta_H = \tan^{-1}\left(\frac{\alpha_H}{\eta_{BS}}\right)\right)$ can vary from 0 to 90°. Generally, for superconductors in the superclean regime, $\theta_H$ is large, which causes $\eta^*$ to be 2-3 orders of magnitude higher than $\eta_{BS}$[89]. However, in this study both NbN and $a$-MoGe the electronic mean free path is much shorter than the coherence length and this mechanism is unlikely to play a significant role.

To understand our results, we note that the low excitation frequency in our experiments affect the vortex dynamics in two different ways. First, at microwave frequencies, vortices get an extremely small time to traverse during every single half-cycle of the ac drive, which gives them very little room for thermal hopping. On the other hand, thermally activated motion across pinning sites becomes more significant over the large time-period at kHz frequencies. Secondly, for an overdamped harmonic oscillator (such as in eqn. (1)), the maximum velocity of the particle increases monotonically with frequency and reaches a limiting value above the depinning frequency, $\alpha_L/\eta$. Thus, in general at kHz frequencies the maximum velocity of the vortex will be much smaller than that for GHz frequencies for the same drive current. This is expected to have two effects. First, the interaction of the vortices with the random pinning potential becomes important[100,101]. It has been shown that due to the interaction with random pinning the viscosity acquires an additive term such that, $\eta = \eta_{BS} + \gamma(v)$, where $\gamma(v)$ is the contribution to the viscous drag from vortex-pin interactions that depends on the velocity ($v$) of the vortex as $\gamma(v) \sim v^{-1/2}$ [100]. At present there is no good theoretical estimate of the magnitude of $\gamma(v)$; it is expected to have a temperature dependence of the form $\gamma \propto e^{U/k_BT}$; consequently, this factor can become very large when the vortex velocity is small, particularly at low temperatures. The second effect is the viscosity arising from the shear deformation of the vortex lattice under an ac drive. Carruzzo *et al.*[102] showed that at low frequencies vortex-vortex interactions lead to a shear flow of vortices when some vortices shove past others under the influence of an ac drive. For a 2D vortex lattice, the drag associated with shear modifies the BS term, introducing a multiplicative factor[103] $e^{U_{shear}/k_BT}$, where $U_{shear} \sim c_{66}V_c$ [102], where $c_{66}$ is the shear modulus of the vortex lattice and $V_c$ is the volume change due to shoving and rearrangement of vortices. Both these effects can lead to several orders of magnitude enhancement in $\eta$ over the BS value, even though it is difficult to distinguish the relative contribution of these two effects. Both these effects provide plausible explanation for the large viscosity observed in our experiments. However, we would like to note that we do not observe the activated temperature dependence of $\eta$ predicted in these theories. However, the activated form for the temperature dependence which implies a diverging $\eta$ as $T\rightarrow0$ is likely to be a first approximation and phenomenological modifications to this form have also been proposed in literature[104]. Therefore, further theoretical studies are needed to understand the temperature dependence of $\eta$.

## 6. Conclusion

We have investigated the complex screening response at kHz frequencies in the mixed state of two superconducting films that are in two extremes in terms of pinning strength: strongly pinned NbN thin film and very weakly pinned *a*-MoGe thin film. Our results show that in both systems the low frequency vortex response is characterised by a vortex viscosity that exceed the BS estimate by several orders of magnitude. While vortex-pin and vortex-vortex interactions could account for such an enhancement at present our data is not consistent with the temperature dependence proposed for these effects. We have also shown that using phenomenological temperature dependence of pinning parameters we can reasonably capture the temperature dependence of the shielding response in the vortex state. This calls for further theoretical studies to obtain a complete description of the low frequency vortex response in superconducting thin films.

## 7. Acknowledgement

This work was financially supported by Department of Atomic Energy, Government of India. We thank Ganesh Jangam for help with magnetisation measurements.

## 8. Author Contribution

SB and SM performed the screening response measurements and SB analysed the data. Sample growth and basic characterisation was done by SB, SM, VB and JJ. PR conceived the problem and supervised the project. The paper was written by SB and PR. All authors commented on the manuscript.

## Appendix: Fitting procedure and estimation of goodness of fit

We fit the temperature variation of $\lambda^{-2}$ and $\delta^{-2}$ at different magnetic fields with the CC model as follows. To the data on NbN, we start with eqn. (12) and vary the four parameters $T_0, U_0, \alpha_{L0}$ and $\eta_0$ to obtain the best fit for $\lambda^{-2}$ by maximizing $R^2$. The variation is done in such a way, that the values for $\alpha_L(T)$ and $\eta(T)$ at 2K match closely with the values extracted using the GR model from the raw data. Ideally the values for $\alpha_L(T)$ and $\eta(T)$ at low temperatures (much below $T_C$,) should be the same in both the CC and the GR model, as flux creep being negligible, it reduces the CC model to the GR model. $R^2$ is defined as $1 - \frac{SS_{res}}{SS_{tot}}$, where the residual sum, $SS_{res} = \sum_i (y_i - f_i)^2$, $y_i$ denoting the experimental data at a point $i$ and $f_i$ denotes the simulated value at the same point; $SS_{tot}$ is defined as $\sum_i (y_i - \bar{y})^2$ where $\bar{y}$ is the arithmetic mean of the experimental data. For a good fit, $R^2 \sim 1$. To maintain an optimal fit, we have chosen the lower bound of $R^2$ estimate as $R^2 > 0.99$ for $\lambda^{-2}$ and $R^2 > 0.84$ for $\delta^{-2}$ and hence small deviations in the values of $\alpha_L(T)$ and $\eta(T)$ can be seen in fig. 6 (b) and 6 (c).

In the case of *a*-MoGe, we adopt a similar procedure but with a small difference. Here we primarily maximise $R^2$ for $\lambda^{-2}$ while at the same time try to reproduce the qualitative variation of $\delta^{-2}$. This is because within our model we cannot quantitatively capture the magnitude of the dissipation peak which is larger in our simulations than in experiments. This deviation is mainly caused by our assumed temperature dependence of different parameters. In principle, one can improve the agreement by fine tuning the temperature dependence of different parameters. However, this will necessarily introduce more parameters. Since this does not provide us with any additional physical insight, we refrain from doing so in this paper.